\documentstyle[jkas32]{article}

\def\fdg{\hbox{$.\!\!^\circ$}}
\def\hdg{\hbox{$.\!\! \rm ^h$}}
\def\sdg{\hbox{$.\!\! \rm ^s$}}

\runningauthor{LEE ET AL.}
\runningtitle{INFRARED EXCESS AND MOLECULAR GAS}
\beginpage{41}
\endpage{53}

\begin{document}

\title{INFRARED EXCESS AND MOLECULAR GAS IN GALACTIC SUPERSHELLS}
\author{
{\large\textbf{\textsc{J}}}\textbf{\textsc{EONG}}$-${\large\textbf{\textsc{E}}}\textbf{\textsc{UN}}
{\large\textbf{\textsc{L}}}\textbf{\textsc{EE}},
{\large\textbf{\textsc{K}}}\textbf{\textsc{EE}}$-${\large\textbf{\textsc{T}}}\textbf{\textsc{AE}}
{\large\textbf{\textsc{K}}}\textbf{\textsc{IM}},
\textbf{\textsc{AND}}
{\large\textbf{\textsc{B}}}\textbf{\textsc{ON}}$-${\large\textbf{\textsc{C}}}\textbf{\textsc{HUL}}
{\large\textbf{\textsc{K}}}\textbf{\textsc{OO}}
}

\address{Department of Astromony, Seoul National University, 
   Seoul 151-742, Korea \\
{\it E-mail: jelee@astro.snu.ac.kr}
}
\address{\normalsize{\it (Received Mar. 19, 1999; Accepted Apr. 9, 1999)}}

\abstract{
We have carried out high-resolution observations along one-dimensional cuts
through the three Galactic supershells GS 064$-$01$-$97, GS 090$-$28$-$17, 
and GS 174$+$02$-$64 in the HI 21 cm and CO J=1$-$0 lines. 
By comparing the HI data with IRAS data, we have derived the distributions 
of the $I_{100}$ and $\tau_{100}$ excesses, which are, respectively,
the 100 $\rm \mu m$ intensity and 100 $\rm \mu m$ optical depth in excess 
of what would be expected from HI emission.
We have found that both the $I_{100}$ and $\tau_{100}$ excesses have good
correlations with the CO integrated intensity $W_{\rm CO}$ in all three 
supershells. But the $I_{100}$ excess appears to underestimate $\rm H_2$
column density $N({\rm H_2})$ by factors of 1.5$-$3.8.
This factor is the ratio of atomic to molecular infrared emissivities,
and we show that it can be roughly determined from the HI and IRAS data.
By comparing the $\tau_{100}$ excess with 
$W_{\rm CO}$, we derive the conversion factor 
$X \equiv N({\rm H_2})/W_{\rm CO}\simeq 0.26-0.66$ in the three supershells. 
In GS 090$-$28$-$17, which is a very
diffuse shell, our result suggests that the region with $N({\rm H_2})\lesssim
3\times10^{20}~\rm cm^{-2}$ does not have observable CO emission, which
appears to be consistent with previous results indicating that 
diffuse molecular gas is not observable in CO.
Our results show that the molecular gas has a 60/100 $\rm \mu m$ color
temperature $T_{\rm d}$ lower than the atomic gas. The low value of $T_{\rm d}$
might be due either to the low equilibrium temperature or to the lower
abundance of small grains, or a combination of both. }


\maketitle

\begin{table*}
\begin{center}
{\bf Table 1.} ~Parameters of HI supershells\\
\vskip 0.3cm
\begin{tabular}[t]{cccccccc}
\hline\hline
 Name & $\Delta l$ & $\Delta b$ & $\rm V_{min}$ & $\rm V_{max}$ & d
& $\rm V_{sh}$ & Reference\\
 &(degree) & (degree) & ($\rm km~s^{-1}$) & ($\rm km~s^{-1}$) & (kpc)
&($\rm km~s^{-1}$)  & \\
\hline
 GS 064$-$01$-$97 & 11 & 6 & $-99$ & $-75$ & 16.9 & 22 & 1\\
 GS 090$-$28$-$17 & 18 & 21 & $-22$ & $-13$ & 3.8 & ... & 2\\
 GS 174$+$02$-$64 & 38 & 28 & $-90$ & $-38$ & ... & ... & 2\\
\hline
\end{tabular}
\end{center}
\vskip -0.2cm
~~~~~~~~~~~~~~~~~~~~~~RERERENCES.$-$(1) Heiles (1979); (2) Heiles (1984).\\
\end{table*}

\section{INTRODUCTION}

Since molecular hydrogen ($\rm H_2$), which constitutes the majority of
interstellar molecules, has a large excitation temperature and 
no permanent electric dipole moment, its emission is not usually observable in
a cold interstellar medium. Instead, the emission lines from 
the rotational transitions of CO have become the most widely used probe of the
interstellar molecular gas. 
There are, however, uncertainties in using CO emission lines
because the fractional abundance of CO may vary from cloud to
cloud, sometimes even within a cloud, and the line intensity 
depends on various excitation conditions.

The {\it InfraRed Astronomical Satellite} (IRAS) (Low et al. 1984) 
has provided  a new method to study interstellar molecular gas.
Since the interstellar gas content can be estimated from the dust content,  
several groups have proposed the infrared (IR) excess, 
which is the 100 $\rm {\mu m}$ intensity
in excess of what would be expected from HI emission,
as an indicator of $\rm H_2$ column density in regions where 
the amount of ionized gas is negligible. 
This idea is based on the assumption that far-infrared emission arises from 
dust grains well-mixed with interstellar gas (Mathis, Mezger, \& Panagia
1983) and the observational result that the 100 $\rm {\mu m}$ intensity 
is tightly correlated with HI column density (Low et al. 1984; 
Boulanger et al. 1985; de Vries, Heithausen, Thaddeus 1987;
Boulanger \& Perault  1988; D\'{e}sert, Bazell, \& Boulanger 1988 [DBB]; 
Heiles, Reach, \& Koo 1988).
However, CO observational studies on the ``IR-excess'' clouds showed that 
there is a substantial difference between the distribution of IR excess 
and that of CO emission.
Using the IRAS 100 $\rm {\mu m}$ and the Berkeley/Parkes HI
low-resolution ($\sim30'$) all-sky maps, for example, D\'{e}sert et al. (1988) 
investigated the global distribution of IR-excess clouds 
at $\rm |b|\geq 5^\circ$.  They found corresponding entries in their 
catalog for only 48\% of the high-latitude clouds detected by 
Magnani, Blitz, \& Mundy (1985). A subsequent CO line survey
has also shown that CO emission was detected in only 13\% (27/201) 
of the IR-excess clouds in the DBB catalog (Blitz et al. 1990). 
Blitz et al. (1990) suggested that CO does not trace all of the molecular 
gas due to the low abundance, $\rm [CO]/[H_2] \sim 10^{-6}$, 
in the remaining clouds.  Heithausen et al. (1993) found in the North pole 
region that only 26\% of IR-excess clouds in the DBB catalog have detectable 
CO and that $\sim 2/3$ of the molecular clouds with CO do not have an IR excess.
By comparing a high resolution ($1'$$-$$2'$) CO map with an IR-excess map for 
an isolated interstellar cirrus cloud, HRK236+39, Reach et al. (1994) 
showed that the IR-excess regions are much larger than the CO-emitting regions 
and that the CO peaks correspond to local maxima of the IR excess. 
Meyerderks \& Heithausen (1996) also found that the distributions of 
the IR excess and CO did not match in the Polaris flare region. 

In Paper I (Kim, Lee, \& Koo 1999), we showed that the discrepancy between 
the distribution of the $I_{100}$ excess and that of CO emission is due to 
two factors: (1) enhanced 100 $\rm {\mu m}$ intensity without detectable CO, and
(2) the low infrared emissivity of molecular clouds.
We compared the $I_{100}$ excess with CO emission in the Galactic worm 
GW46.4+5.5, which is a long filamentary structure extending vertically 
from the Galactic plane.
In GW46.4+5.5, we found that the enhanced heating 
by a massive star produces {\it $I_{100}$ excess without CO emission}, 
while the low infrared emissivity of dust grains associated with molecular gas 
could completely hide the presence of molecular gas in the infrared.
We showed that the $\tau_{100}$ excess, which is the 100 $\rm {\mu m}$ optical
depth in excess of what would be expected from HI emission, 
could be used as an accurate indicator of molecular content along a line of
sight. Our result suggested that the $I_{100}$ excess could still be 
used to estimate the molecular content if the different emissivities 
of atomic and molecular gases are reflected in the calculation. 
We introduced a correction factor, $\xi_c \equiv <I_{100}/N({\rm H)>_{HI}}
/<I_{100}/N(\rm H)>_{H_2}$ ($\sim2$ in GW46.4+5.5), 
where $<I_{100}/N(\rm H)>_{HI, H_2}$ is the mean 100 $\rm {\mu m}$ 
intensity per unit column density of hydrogen nuclei in atomic
or molecular gases.

In this paper, we study the IR excess and molecular gas in Galactic 
supershells, which are large shell-like structures of interstellar HI 
gas that appear to be either expanding or stationary (Heiles 1979, 1984).
As in Paper I, we have carried out high-resolution observations 
along one-dimensional cuts in the HI 21 cm and CO J=1$-$0 lines 
through Galactic supershells and made a comparison of 100 
$\rm {\mu m}$ excess and CO content.
While Paper I dealt with only a small region at $l=\rm 46^\circ$ near 
the galactic plane($|b|\leq \rm 5^\circ$), here we study several areas 
at different galactic longitudes and latitudes.
In Section II we describe the HI and CO observations. 
In Section III we evaluate the $I_{100}$ and $\tau_{100}$ excesses 
using {\it IRAS} and HI data, and compare their distributions with
that of the integrated intensity of CO emission. 
We discuss our results in Section IV, and summarize the paper in Section V.

\section{OBSERVATIONS}

HI 21-cm line observations were made using the 305-m telescope 
(HPBW~$\sim3.3'$) at Arecibo Observatory, in 1990 October.
Both circular polarizations were observed simultaneously using two 1024-channel
correlators, with a total bandwidth of 5 MHz each, so that the velocity
resolution was 2.06 $\rm km~s^{-1}$ after Hanning smoothing.
Each spectrum was obtained by integrating for one minute using 
frequency switching.
The antenna temperature was converted to the brightness temperature using
a beam efficiency of 0.84, which was estimated previously (Koo et al. 1990).
We observed three supershells: GS 064$-$01$-$97, GS 090$-$28$-$17, 
and GS 174$+$02$-$64
(hereafter GS064, GS090, GS174) in the catalogs of Heiles (1979, 1984).
These regions are appropriate for a detailed study of the correlation
between the gas content and infrared emission
because they include distinct filamentary structures.
Table 1 lists their parameters. In Table 1, column 1 contains the shell name,
which specifies the approximate longitude, latitude, and LSR velocity of the
shell center. Columns 2 and 3 contain the shell diameter in longitude and
latitude, respectively. Columns 4 and 5 contain the approximate minimum and 
maximum LSR velocities at which the shell is visible.
Column 6 contains the distance of the shell from the sun, and column 7 contains
the expansion velocity of the shell. 
Figure 1 shows the infrared morphology of the parts of the supershells 
related to this work.
GS064 is close to the galactic plane and is not clearly discernible in Figure 1.
The typical HI column density and 100 $\rm {\mu m}$ intensity are 
$N(\rm HI)\simeq5-9\times10^{21}~\rm cm^{-2}$ and 
$I_{100}\simeq50-140~\rm MJy~sr^{-1}$.
GS090 is located at high latitude and is very diffuse.  
Typically, $N(\rm HI)\simeq3-7\times10^{20}~\rm cm^{-2}$ and 
$I_{100}\simeq2-10~\rm MJy~sr^{-1}$. 
GS174 has a very large angular size and is known as an anticenter shell.
Typically, $N(\rm HI)\simeq1-5\times10^{21}~\rm cm^{-2}$ and 
$I_{100}\simeq10-50~\rm MJy~sr^{-1}$.
The regions observed in HI 21-cm line are listed in Table 2 and they are marked 
as solid lines in Figure 1. 
Spectra were sampled at every $3'$ or $4'$.
\begin{figure*}
\vspace{20cm}
\caption{\label{fig:fig1ab} 
100 $\rm \mu m$ intensity maps of the Galactic Supershells
studied in this paper. The solid lines indicate one-dimensional cuts along
which HI 21 cm line observations were made.
CO observations were made toward IR-excess regions in these one-dimensional
cuts.
(a) GS 064$-$01$-$97.
Grey-scale varies from 50 to 170 $\rm MJy~sr^{-1}$, and contour levels are
60, 80, 100, 120, 140, and 160 $\rm MJy~sr^{-1}$.
(b) GS 090$-$28$-$17. Grey-scale varies from 1 to 14 $\rm MJy~sr^{-1}$,
and contour levels are 3.5, 5.5, 7.5, 9.5, 11.5, and 13.5 $\rm MJy~sr^{-1}$.
(c) GS 174$+$02$-$64. Grey-scale varies from 15 to 50
$\rm MJy~sr^{-1}$, and contour levels are
20, 25, 30, 35, and 40 $\rm MJy~sr^{-1}$.
}
\includegraphics{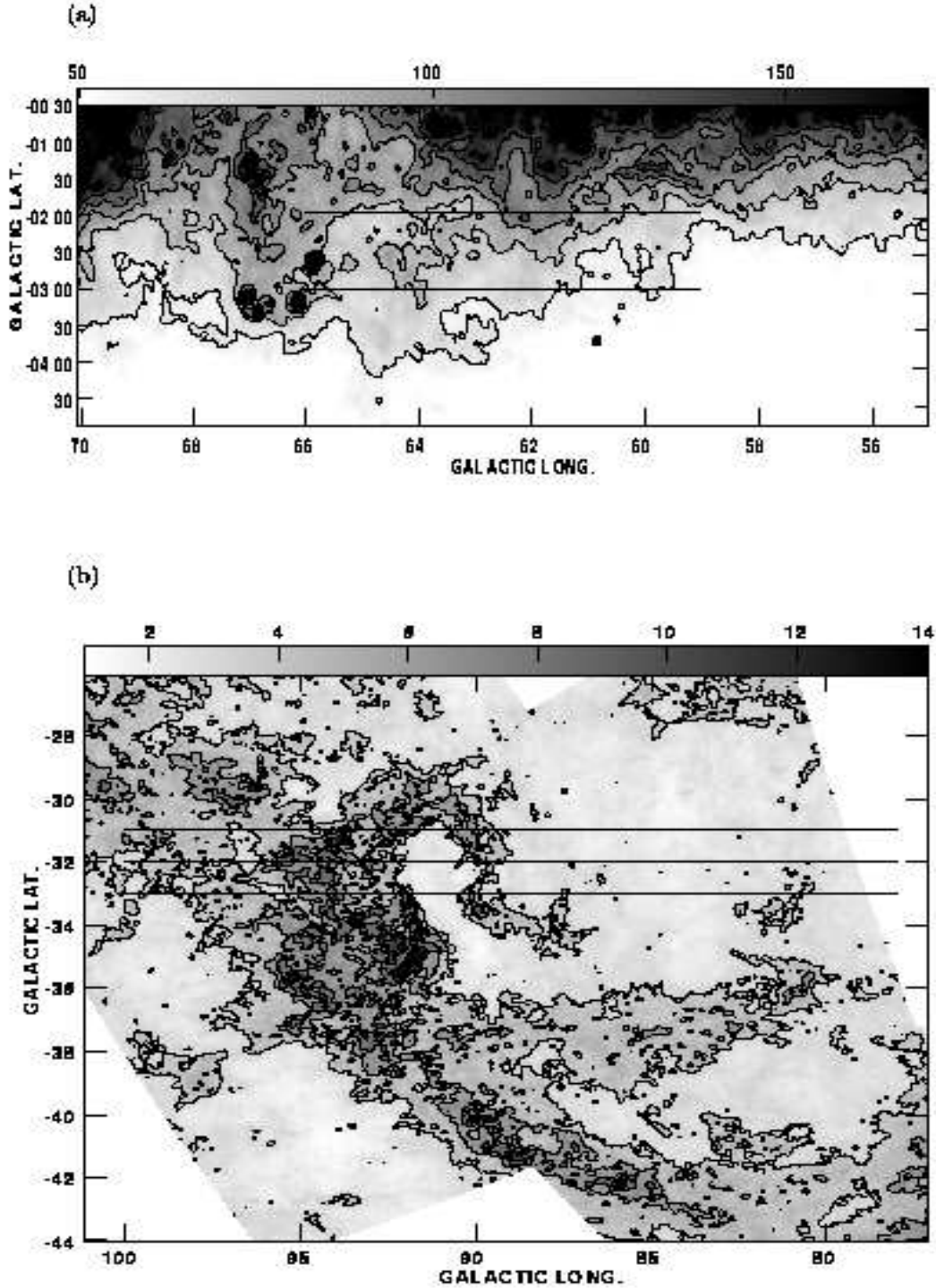}
\end{figure*}
 
\begin{figure*}
\vspace{10cm}
\begin{center}
{\bf Fig. 1.$-$} 
Continued
\end{center}
\includegraphics{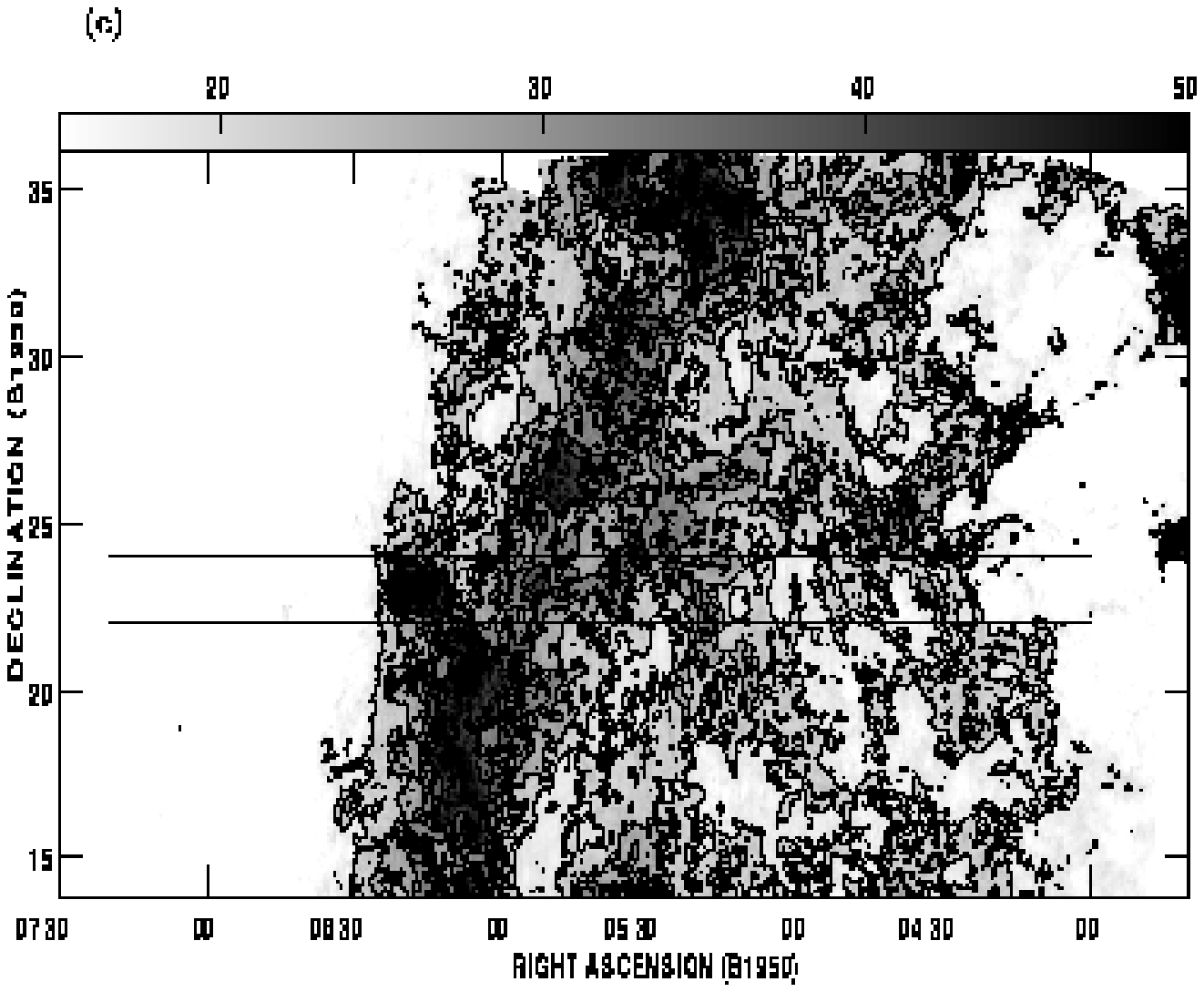}
\end{figure*}

CO J=1$-$0 line observations were made using
the 14-m telescope (HPBW~$\sim50''$) at Taeduk Radio Astronomy Observatory 
(TRAO) during 1996 March, May, November, and 1997 May.
The data were collected using a 256-channel filter bank with a total
bandwidth of 64 MHz. The velocity coverage was from $-80$ to $+80\rm~km~s^{-1}$,
and the velocity resolution was 0.65 $\rm km~s^{-1}$. 
The system temperature varied in the range 700$-$1000 K 
during the observing sessions depending on weather conditions and elevation of 
the source. 
The typical rms noise of the spectra was 0.1$-$0.2 K.
The antenna temperature was converted to the brightness temperature 
using a main beam efficiency of 0.39.
We first observed IR-excess peaks brighter than
3 MJy $\rm sr^{-1}$ at 100 $\rm {\mu m}$, which is the brightness corresponding 
to a CO J=1$-$0 line intensity of $\sim0.1$ K. 
For the peak positions with detectable CO emission, we have made further 
observations along the one-dimensional cuts where HI observations had been made.
CO spectra were obtained at every $3'$ or $4'$ using position switching.
Observed regions and reference positions are listed in Table 2.
The reference positions were checked to have no appreciable ($ < 0.1$ K) 
CO J=1$-$0 emission.

\section{INFRARED EXCESS AND MOLECULAR GAS}
\subsection{HI gas and Infrared excess} 

Figure 2 shows some representative HI and CO spectra of each object.
HI spectra usually have several velocity components, and we integrated over 
the whole velocity range to derive $N(\rm HI)$ along the line of sight.
Assuming that the 21-cm line emission is optically thin, we converted the 
observed brightness temperature to $N(\rm HI)$.

The method of the analysis was identical to that used in Paper I. 
First, in each area, we derived the correlation between 
$N(\rm HI)$ and $I_{100}$, i.e., $I_{100}=a~N({\rm HI})+b$.
In contrast to Paper I, however, we could not exclude all of the points with
detectable CO emission in advance because we did not carry out CO observations
toward the whole area.
In order to exclude the points with possible molecular gas, therefore, 
we repeated least-squares fits using only pixels within $\pm 2\sigma$ 
deviations from the previous fit until the standard deviation did not change.
The parameters $a$ and $b$ determined in that way are listed in Table 3.
The $I_{100}/N(\rm HI)$ ratios of GS064 and GS090 are consistent with values
in the solar neighborhood, while the ratio for GS174 is significantly smaller.
Based on the derived relation between $I_{100}$ and $N(\rm HI)$,
we can determine the $I_{100}$ excess, i.e., the $I_{100}$ 
intensity in excess of what would be expected from $N(\rm HI)$. 
This $I_{100}$ excess can be used to estimate the $\rm H_2$ column density;

\begin{equation}
N({\rm H_2})_{I_{100}}=\frac{1}{2}\left[\frac{I_{100,c}}{<I_{100,c}
/N({\rm H})>}-N(\rm HI) \right]~~~~~(\rm cm^{-2})
\end{equation}

{\noindent
where the subscript `c' indicates the quantity corrected for offset and 
the angle bracket indicates an average ratio.}
In this process, we assumed that the dust-to-gas ratios and infrared 
emissivities of the HI cloud and $\rm H_2$ cloud are same. 
Therefore $<I_{100}/N(\rm H)>$ is equal to $<I_{100}/N(\rm HI)>$.

 \begin{table*}
\vspace{-0.5cm}
 \begin{center}
 {\bf Table 2.} ~Positions of scans and OFF positions\\
 \vskip 0.3cm
 \begin{tabular}[t]{ccccc}
 \hline\hline
 ~ & ~& \multispan{2} $l/\alpha_{1950}$-range 
 & OFFs\tablenotemark{a}\\ \cline{3-4}
 Name & $b/\delta_{1950}$ & HI 21 cm line & CO J=1$-$0 line
 & ($l,b$)/($\alpha,\delta)_{1950}$\\
 \hline
 GS 064$-$01$-$97 & & & &\\ \cline{1-1}
 & $-$3\fdg00 & 66\fdg00$-$59\fdg00
 &59\fdg75$-$59\fdg45 & (59\fdg00,$-$3\fdg00)\\
 & & &66.00$-$61.55 & (60.00,$-3.50$),(63.00,$-4.00$),(65.50,$-4.00$)\\
 &$-2.00$ & 66.00$-$59.00 & 59.65$-$59.00 & (59.00,$-2.50$)\\
 & & & 62.75$-$60.02 & (59.00,$-2.50$)\\
 & & & 65.30$-$63.85 & (64.70,$-3.30$)\\
 GS 090$-$28$-$17 & & & & \\ \cline{1-1}\\
 &$-$31\fdg00 & 100\fdg0$-$78\fdg00
 &94\fdg67$-$91\fdg60
 &(92\fdg00, $-$31\fdg50),(94.00,$-$30.60)\\
 & $-32.00$ & 100.0$-$78.00 & 94.50$-$92.47 & (93.00,$-31.50$)\\
 & $-33.00$ & 100.0$-$78.00 & 94.33$-$91.93
 & (91.00, $-33.00$),(95.50,$-33.00$)\\
 GS 174$+$02$-$64 & & & & \\ \cline{1-1}\\
 &~~22\fdg00 & ${\rm 4^h00^m00^s}$
 & ${\rm 4^h07^m48^s}$$-$${\rm 4^h05^m24^s}$
 &(${\rm 4^h04^m12^s}$, 21\fdg00)\\
& & ~~$-$${\rm 7^h20^m00^s}$& 4~33~00$-$4~19~12 
& (4~15~00,18.50),(4~30~00,21.50)\\
 & & & 4~45~00$-$4~41~24 & (4~43~12,21.50)\\
 & & & 5~21~00$-$5~15~00 & (5~18~00,21.00)\\
 & & & 6~18~00$-$5~51~00 & (5~30~00,22.00),(6~15~36,21.00)\\
 &~~24.00 & 4~00~00  & 4~34~48$-$4~06~00
 & (4~04~48,21.00),(4~04~48,23.50),(4~45~00,23.00)\\
& & ~~$-$7~20~00 &  5~42~00$-$5~30~00 & (5~36~00,25.00)\\
 & & & 6~04~48$-$5~52~12 & (6~00~00,25.00)\\
 \hline
 \end{tabular}
 \end{center}
 \vskip -0.2cm
 ~~~~~~~~~~~ $^a$OFF positions for CO J=1$-$0 line observations\\
 \end{table*}

\begin{figure*}
\vspace{10.5cm}
\caption{\label{fig:fig2}
Representative HI 21 cm (left column)
and CO J=1$-$0 (right column) spectra of each object.
These are the spectra of the regions where the strongest CO line emission
has been detected. The positions are labeled on the upper left part of
each picture.
}
\includegraphics{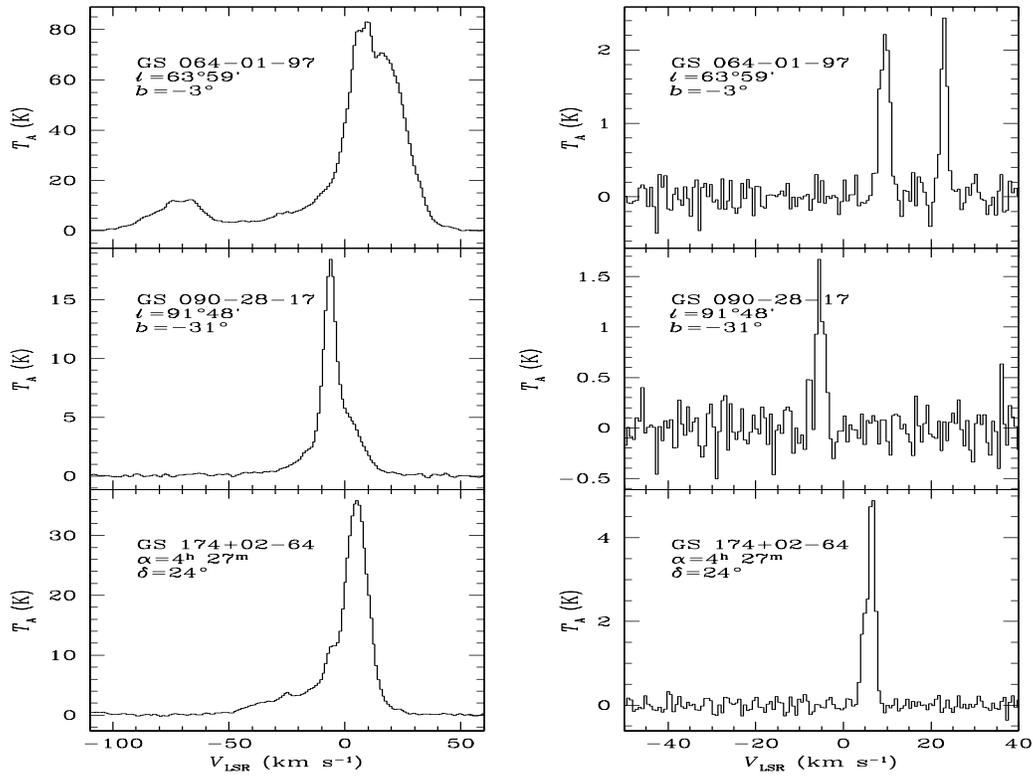}
\end{figure*}

\begin{table*}
\begin{center}
{\bf Table 3.} ~The correlation between $I_{100}$ and $N(\rm HI)$\\
\vskip 0.3cm
\begin{tabular}[t]{ccc}
\hline\hline
Name & a  & b \\
&($\rm MJy~sr^{-1}~(10^{20}~cm^{-2})^{-1}$) & ($\rm MJy~sr^{-1}$)\\
\hline
GS 064$-$01$-$97 & 1.30 $\pm$ 0.01 & $-8.15 \pm$0.67 \\
GS 090$-$28$-$17 & 1.05 $\pm$ 0.02 & $-2.00 \pm$0.07 \\
GS 174$+$02$-$64 & 0.40 $\pm$ 0.01 & 12.13 $\pm$0.18 \\
\hline
\end{tabular}
\end{center}
\vskip -0.2cm
~~~~~~~~~~~~~~~~~~~~~~~~~~~~~~~~~~~~~~~~~Note : $I_{100} ~~=~~ a N({\rm HI}) ~+~
b $ \\
 \end{table*}

We may alternatively use the 100 $\rm \mu m$ optical depth $\tau_{100}$ for
identifying the region with molecular gas. Using 60 and 100
$\rm \mu m$ intensities, we derived the color temperature, $T_{\rm d}$, and 
$\tau_{100}$. In each area, we determined a mean relation between $\tau_{100}$
and $N(\rm HI)$, and derived the $\tau_{100}$ excess, i.e., 
the 100 $\rm \mu m$ optical depth in excess of what would be expected from
$N(\rm HI)$. This $\tau_{100}$ excess can be used to estimate a separate 
$\rm H_2$ column density;

\begin{equation}
N({\rm H_2})_{\tau_{100}}=\frac{1}{2}\left[\frac{\tau_{100,c}}
{<\tau_{100,c}/N({\rm H})>}-N(\rm HI)\right]~~~~~(\rm cm^{-2})
\end{equation}

Figures 3$-$5 show the distributions of the derived  $T_{\rm d}$, 
$N({\rm H_2})_{I_{100}}$, and $N({\rm H_2})_{\tau_{100}}$ 
in the three supershells. 
In GS064, the distributions of $N({\rm H_2})_{I_{100}}$ and 
$N({\rm H_2})_{\tau_{100}}$ are very similar except at several positions 
where $N({\rm H_2})_{\tau_{100}}$ produces a sharp minimum, e.g., at
$(l,b)=($61\fdg3, $-$3\fdg0) and (59\fdg95, $-$2\fdg0).
These are the directions toward IRAS point sources with large 60 $\rm \mu m$
fluxes (see Table 4), so that the derived $T_{\rm d}$'s appear sharply peaked.
In GS090, the distributions of $N({\rm H_2})_{I_{100}}$ and 
$N({\rm H_2})_{\tau_{100}}$ appear similar too. 
But an important difference is that $N({\rm H_2})_{\tau_{100}}$ is greater 
than $N({\rm H_2})_{I_{100}}$ by a factor of $\sim2.5$. 
This is consistent with the result of Paper I showing that the 
{\it $I_{100}$ excess significantly underestimates the amount of 
molecular gas along the line of sight}. 
Another thing to note is that the areas with large $N({\rm H_2})_{\tau_{100}}$
appear to have lower color temperatures, e.g., at 
$(l,b)=(92^\circ-95^\circ$, $-31^\circ$) and ($92^\circ-95^\circ$, $-33^\circ$).
As we show in section III (b), these are the regions with molecular gas. 
In GS174, we again note that $N({\rm H_2})_{\tau_{100}}$ is greater than
$N({\rm H_2})_{I_{100}}$, particularly at 
$(\alpha, \delta)=($4\hdg4$-$4\hdg6, $24^\circ$). 
However, this the factor is found to vary over the region.
There are several positions with very high $T_{\rm d}$'s in GS090 and GS174 
too, and they are listed in Table 4.

$N({\rm H_2})_{I_{100}}$ would significantly underestimate the molecular
content because the IR emissivity of molecular gas, 
$<I_{100}/N({\rm H})>_{\rm H_2}$, could be much lower than that of
atomic gas $<I_{100}/N({\rm H})>_{\rm HI}$.
In Paper I, we introduced a correction factor

\begin{equation}
\xi_c \equiv \frac{<I_{100}/N({\rm H)>_{HI}}}{<I_{100}/N(\rm H)>_{H_2}}
\end{equation}

{\noindent that should be applied to $N({\rm H_2})_{I_{100}}$ in order 
to account for the different IR emissivities of molecular clouds.}
The correction factor could be estimated from 
$N({\rm H_2})_{\tau_{100}}/N({\rm H_2})_{I_{100}}$ and, from Figure 3$-$5,
we find that $\xi_c \simeq 1-1.7$, 2.5, and 3 for GS064, GS090, and GS174.
These correction factors, however, are underestimated because 
$N({\rm H_2})_{\tau_{100}}$ is underestimated. Note that one of our
assumptions in deriving $N({\rm H_2})_{\tau_{100}}$ was a uniform color
temperature for dust along the line of sight. 
But, in general, the 60/100 $\rm \mu m$ color temperature of molecular gas 
is less than that of atomic gas (see section IV).
Therefore, by assuming a uniform color temperature, we overestimate 
$T_{\rm d}$ and underestimate $\tau_{100}$ (and $N({\rm H_2})_{\tau_{100}}$)
of molecular gas. The effect becomes greater as the relative amount of
atomic gas increases (see Fig. 6 of Paper I). Hence, for the supershells
studied in this paper, the effect is largest ($\sim$ 35\%) for GS064
where $N({\rm HI})/N({\rm H_2}) \geq 10$ and the size of this effect
is about 20\% for the other supershells.
If we consider this effect, the true correction factors are 
$\xi_c \simeq 1.5-2.6$, 3.1, and 3.8 for GS064, GS090, and GS174,
respectively.

\begin{figure*}
\vspace{21cm}
\caption{\label{fig:fig3}
 Distributions of (a) $I_{100}$, $N(\rm HI)$, (b) $T_{\rm d}$,
(c) $N({\rm H_2})_{\tau_{100}}$, $N({\rm H_2})_{I_{100}}$,
and (d) $W_{\rm CO}$ for GS 064$-$01$-$97.
In (a), the thin and thick lines represent $N(\rm HI)$ and $I_{100}$,
and, in (c), $N({\rm H_2})_{I_{100}}$ and $N({\rm H_2})_{\tau_{100}}$.
In (d), the thick lines at the bottom indicate the regions
where the CO line observations were carried out.
}
\includegraphics{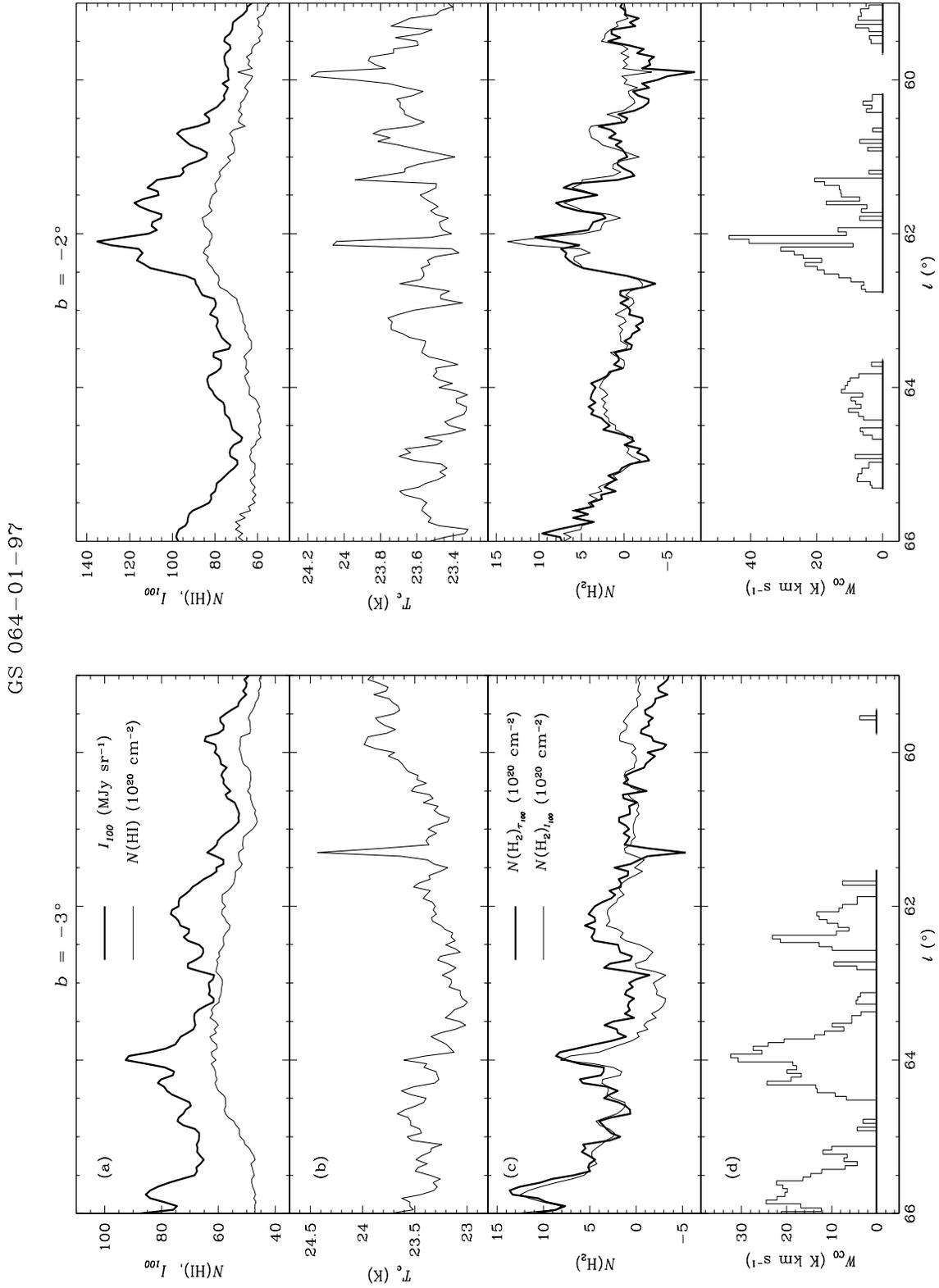}
\end{figure*}

\begin{figure*}
\vspace{22cm}
\begin{center}
{\bf Fig. 4.---}
Same as Fig. 3 but for GS 090$-$28$-$17.
\end{center}
\includegraphics{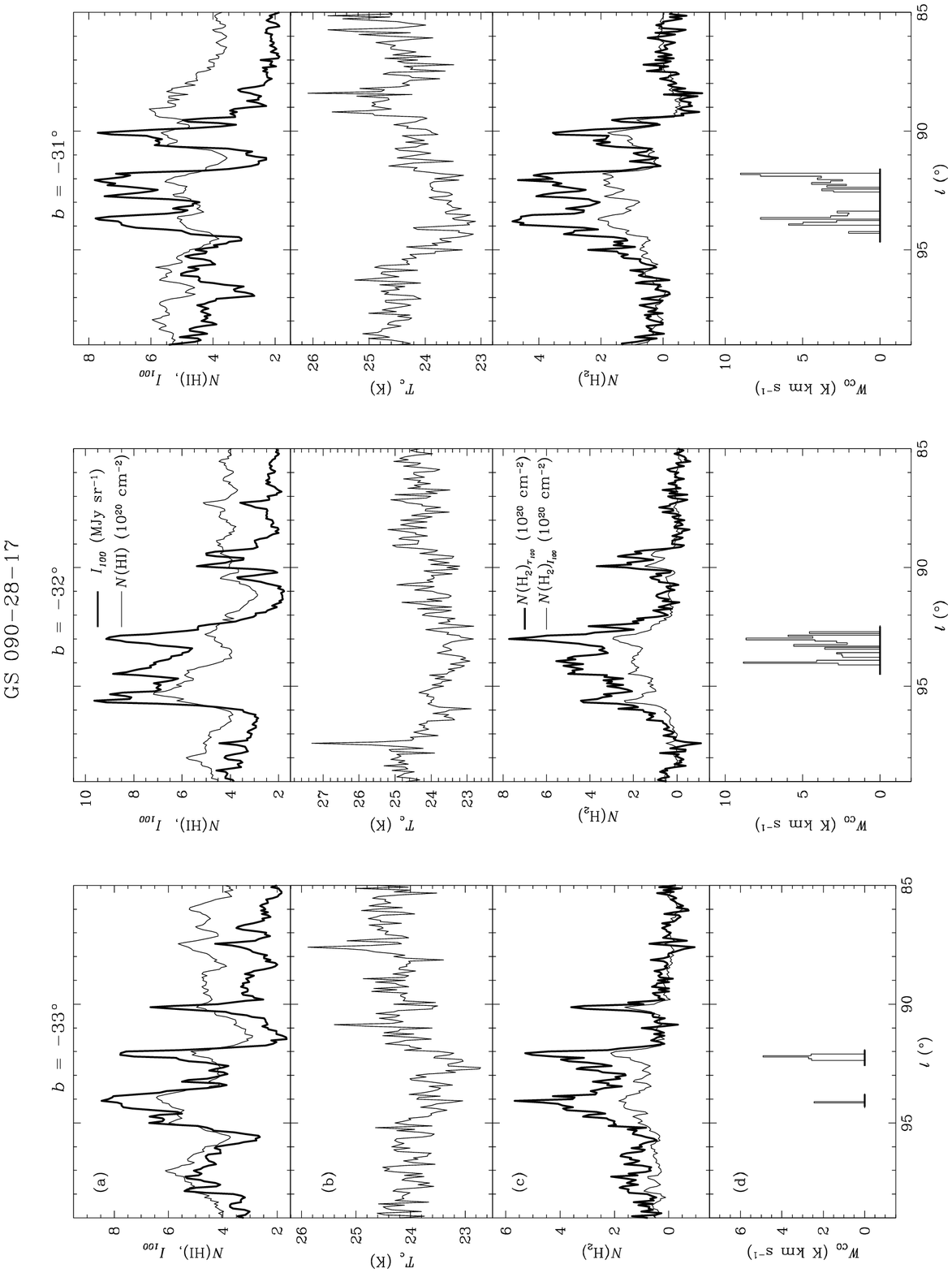}
\end{figure*}

\begin{figure*}
\vspace{21cm}
\begin{center}
{\bf Fig. 5.---}
Same as Fig. 3 but for GS 174$+$02$-$64.
\end{center}
\includegraphics{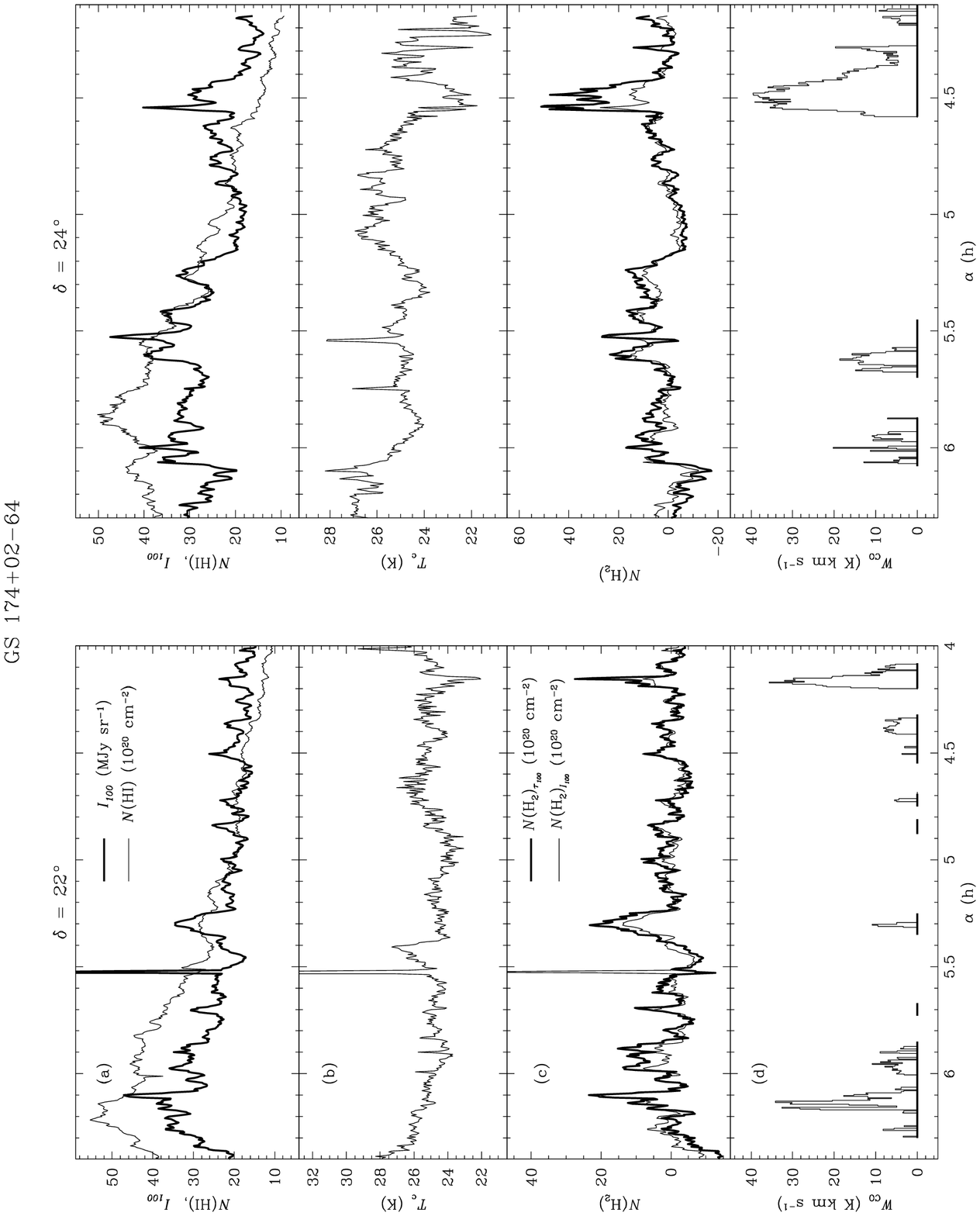}
\end{figure*}

\begin{table*}
\begin{center}
{\bf Table 4.} ~Positions with very high $T_{\rm d}$\\
\vskip 0.15cm
\begin{tabular}[t]{ccccc}
\hline\hline
Name & Positions  & IRAS Point Source & Note & Reference\\
\hline
GS 064$-$01$-$97 &&&&\\ \cline{1-1}\\
&(61\fdg30, $-$3\fdg00)&IRAS 19557+2320&External galaxy&1 \\
&(62\fdg15, $-$2\fdg00)&IRAS 19541+2334&?&1 \\
&(59\fdg95, $-$2\fdg00)& ...    & ?                         & \\
GS 090$-$28$-$17 &&&\\ \cline{1-1}\\
&(87\fdg60, $-$33\fdg00)&  ...    & ?                       & \\
&(97\fdg40, $-$32\fdg00)&IRAS 23129+2548&?&1 \\
GS 174$+$02$-$64 &&&&\\ \cline{1-1}\\
&(${\rm 5^h24^m}$37\sdg8, $21^\circ53'46''$)&... &HD35708 (B2.5IV)& 2\\
&(${\rm 5^h31^m}$31\sdg4, $21^\circ58'54''$)
&IRAS 05315+2158&Crab supernova remnant& 1,4 \\
&(${\rm 4^h32^m}$31\sdg5, $24^\circ02'07''$)&IRAS 04325+2402&
Young stellar object &1,3\\
&(${\rm 5^h32^m}$23\sdg7, $24^\circ00'28''$)&... &HD36819 (B2.5IV)& 2\\
&(${\rm 5^h44^m}$45\sdg5, $24^\circ00'00''$)&IRAS 05447+2400&?&1 \\
\hline
\end{tabular}
\end{center}
\vskip -0.2cm
~~~~~~~~~~~~REFERENCES.$-$(1) IRAS point source catalog (1988);
(2) Gaustad \& Van Buren (1993);

{\indent~~~~~~~~~~~~~~~~~~~~~~~~~~~~~~~~~~~~(3) Kenyon et al. (1993);
(4) Strom \& Greianus (1992).}\\
\end{table*}

\begin{figure*}
\vspace{9cm}
{\bf Fig. 6.---}
$N({\rm H_2})_{\tau_{100}}$ vs. $W_{\rm CO}$ (top) and
$N({\rm H_2})_{I_{100}}$ vs. $W_{\rm CO}$ (bottom) in three supershells
.
The solid line in each frame represents a least-squares fit to the data
.
\includegraphics{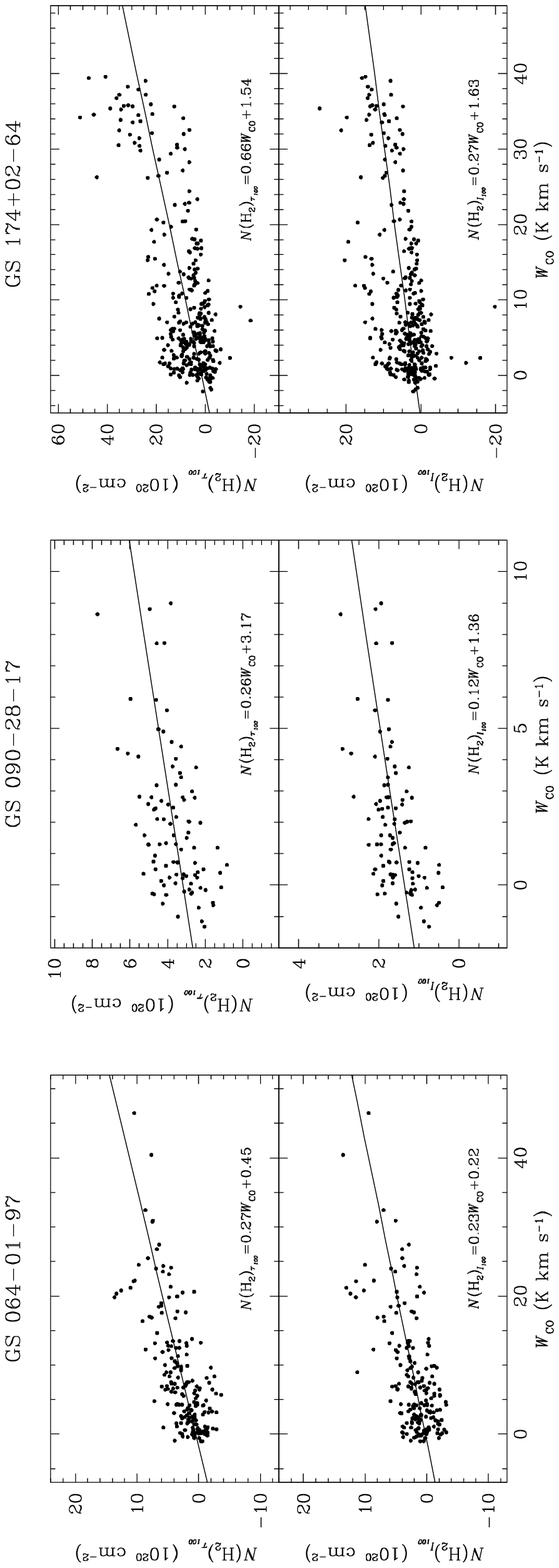}
\end{figure*}

\subsection{Comparison of IR excess and CO emission}

In Figures 3$-$5, we also show the regions where CO observations 
have been carried out and the distributions of the CO integrated 
intensity, $W_{\rm CO}$.
If we compare the distributions of either $N({\rm H_2})_{I_{100}}$ or 
$N({\rm H_2})_{\tau_{100}}$ with those of $W_{\rm CO}$ in Figure 3, 
it is obvious that they are correlated.
The CO detection rate, i.e., the fraction with detectable CO emission among
the observed IR-excess peaks, is very high ($\sim 0.9$) for GS064 
while it is 0.7 for GS090 and GS174. 
$I_{100}$- and $\tau_{100}$-excess peaks have almost the
same detection rate. There are, however, $I_{100}$-excess peaks produced by
local heating sources instead of molecular gas, e.g., 
$(\alpha,\delta)=(\rm 5^h31^m46^s$, $22^\circ$) 
and $(\rm 5^h32^m31^s$, $24^\circ$) in GS174 (see Table 3). 
In such regions, which can be identified by their
high color temperatures, the $I_{100}$-excess cannot trace the molecular
content correctly.

Direct comparison of the $I_{100}$- or $\tau_{100}$-excesses with $W_{\rm CO}$
is not straightforward because the resolution ($50''$) of the CO 
observations is much smaller than those of the HI and IRAS observations.
If $W_{\rm CO}$ varies smoothly, however, the comparison could still be useful.
Figure 6 compares $N({\rm H_2})_{I_{100}}$ and $N({\rm H_2})_{\tau_{100}}$
with $W_{\rm CO}$ in three shells.
In Figure 6, the solid line represents a least-squares fit to the data and
the relation is given in each frame. There are several points to be made
from Figure 6. First, in spite of the different beam sizes, there are
good correlations between $N({\rm H_2})$'s and $W_{\rm CO}$'s.
$N({\rm H_2})_{\tau_{100}}$ and $N({\rm H_2})_{I_{100}}$ have a comparable
degree of correlation with $W_{\rm CO}$. But, as we have pointed out in
section III (a), their conversion factors $X \equiv N({\rm H_2})/W_{\rm CO}$ differ
significantly.  Second, the conversion factor 
varies among supershells, i.e., $X=0.26-0.66$, if we use 
$N({\rm H_2})_{\tau_{100}}$.
This conversion factor is much smaller than the estimated value for
molecular clouds in the Galactic plane, $X=(1.8-4.8)\times10^{20}~\rm cm^{-2}~
(K~ km~s^{-1})^{-1}$ (Scoville \& Sanders 1987), but comparable to that
of high latitude clouds, $X\simeq0.5\times10^{20}~\rm cm^{-2}~
(K~ km~s^{-1})^{-1}$ (de Vries et al. 1987; Heithausen \& Thaddeus 1990; 
Reach et al. 1998).
Some of the molecular gases that we observed in the three supershells 
comprise parts of previously studied molecular clouds. 
In GS064, the observed region is the
western part of the Cygnus Rift (Dame \& Thaddeus 1985). In GS090, it is  
a part of MBM53 (Magnani, Blitz, \& Mundy 1985). In GS174, it covers the 
southern part of the Taurus complex (Ungerechts \& Thaddeus 1987). 
Third, the fit in GS090 has a large offset, e.g., 3.2 $\rm MJy~sr^{-1}$. 
We consider that this is significant. (GS174 also has a large offset. But 
since $N(\rm H_2)$ values are large in GS174, the errors are large. 
Therefore the large offset in GS174 is not significant.) 
The large offset, if it is real,
implies molecular gas without CO emission. There have been suggestions
that diffuse molecular gas would not be observable in CO because CO may
not be shielded from UV photons and because CO may not be collisionally
excited (van Dishoech \& Black 1988; Blitz et al. 1990; Reach et al. 
1994; Meyerdierks \& Heithausen 1996; Boulanger et al. 1998; Reach et al. 1998).
Our result suggests that the critical column density below which the CO line
would not be detected is $\sim 3\times10^{20}~\rm cm^{-2}$. 
This is about half as much as the result of van Dishoech \& Black (1988).

\section{DISCUSSION}

The results of this paper confirm our previous conclusion that 
there are several shortcomings in using the $I_{100}$ excess for estimating
accurate molecular content: First, there are $I_{100}$-excess regions
not due to molecular gas but due to high infrared emissivity, e.g.,
the regions around early-type stars. In this study, the region around
Crab in GS174 was such an example, but it was easily discernible by the large
color temperature. Second, the $I_{100}$ excess would significantly 
underestimate the $\rm H_2$ column density because of the low 
infrared emissivity of molecular gas. In the three supershells that 
we studied, the correction factors are 1.5$-$3.8.

As we have pointed out in Paper I, $\tau_{100}$ and $T_{\rm d}$ are largely
affected by small grains. Previous studies showed that 40$-$60\% of 
60 $\rm \mu m$ emission is contributed by small grains (Sodroski et al. 1994).
Hence, $T_{\rm d}$ could be significantly overestimated compared to
the equilibrium temperature of large grains.
Indeed, this appears to be the case because the derived color temperatures
(23$-$25 K) are significantly larger than the mean dust temperatures of 
the inner (21 K) and outer (17 K) galaxies derived by Sodroski et al. (1994)
using the Diffuse Infrared Background Experiment (DIRBE) 
140 and 240 $\rm \mu m$ observations.
Note that $T_{\rm d}=24$ K when $I_{60}/I_{100}=0.2$ and 
$T_{\rm d}=21$ K where $I_{60}/I_{100}=0.1$ (e.g., see equation (2) of Paper I).
Hence the temperature difference is consistent with the suggestion that
$\sim 50\%$ of 60 $\rm \mu m$ emission is contributed by small grains.
In this case the 100 $\rm \mu m$ optical depth is underestimated by a factor of
$\sim 3$. If the contribution of small dust particles for 60 $\rm \mu m$
emission varies from 10\% to 80\%, this factor varies from 1.2 to 11.

Our results (in this paper and in Paper I) indicate that $T_{\rm d}$ 
derived from $I_{60}/I_{100}$ is low in molecular gas. 
Similar phenomena have been observed in other clouds too
(Boulanger et al. 1990; Laureijs et al. 1991; Bernard et al. 1993).
$T_{\rm d}$ could be low either because the
equilibrium temperature is low or because the abundance of small grains 
is less, or both. Indeed, the equilibrium temperature of dust grains 
in molecular gas is lower than that in atomic gas by 2 K (Sodroski et al. 1994).
Also, in nearby molecular clouds, it was found that the abundance of
small grains decreases in the central regions of molecular clouds
(Bernard et al. 1993).
Boulanger et al. (1990) suggested that the abundance variation occurs
because small grains condense onto grains and do not form
in regions of low color ratio, while they are formed and detached from
grain mantles in regions of high color ratio.

There have been other suggested methods for estimating the molecular content
from $I_{60}$ and $I_{100}$.
Laureijs et al. (1991) showed that $\Delta I_{100} \equiv I_{100}
-I_{60}/\Theta$, where $\Theta$ is the average value for
$I_{60}/I_{100}$ in the outer diffuse region, is strongly correlated
with molecular material with densities of $n_{H_2} > 10^3 {\rm cm^{-1}}$.
Abergel et al. (1994) also found in the Taurus complex that 
$\Delta I_{100}$ is tightly correlated with $W_{\rm {}^{13}CO}$, 
and that $\rm {}^{13}CO$-emitting regions coincide with cold regions traced
by $\Delta I_{100}$. The DIRBE emission at 140 and 240 $\rm \mu m$
could be also used in calculating $\Delta I_\lambda$ (Lagache et al. 1998).
These $\Delta I_\lambda$ measurements seem to be other good indicators of 
molecular gas in dense regions. This is probably
because $I_{60}/I_{100}$ drops in the molecular cloud due to the combined 
effect of the decrease in dust temperature and the change in dust
properties. Boulanger et al. (1998) introduced $\Delta I_{100}$
multiplied by a scaling factor $\Theta/(\Theta-\Theta_{\rm cold})$, where
$\Theta_{\rm cold}$ is the average value for the cold component, to correct
for the fraction of 100 $\rm \mu m$ emission from the cold component
which is lost in the subtraction. The advantage of this method is that
one can search for dense molecular clouds without HI data. We mentioned
above the contribution of small grains to $I_{60}$ as major
problems in applying the $\tau_{100}$-excess method. 
On large angular scales, we may use the DIRBE data for the 100, 140, 
and 240 $\rm \mu m$ wavebands to resolve such problems. 
In that case, $\tau_{140}$ excess and $\tau_{240}$ 
excess are recommended as other indicators of molecular gas.

\section{CONCLUSION}

In this paper, we have confirmed our previous results (Kim et al. 1999) 
suggesting that there are several shortcomings in using the $I_{100}$ 
excess for estimating accurate molecular content: 
First there are $I_{100}$-excess regions
due to high infrared emissivity, and, second the $I_{100}$ excess would
significantly underestimate the $\rm H_2$ column density $N(\rm H_2)$
because of the low infrared emissivity of molecular gas.
We may still estimate $N(\rm H_2)$ from the $I_{100}$ excess by applying the
correction factor $\xi_c=<I_{100}/N({\rm H)>_{HI}}/<I_{100}/N(\rm H)>_{H_2}$,
which could be roughly determined from HI and IRAS data. Our results 
show that $\xi_c$ varies from region to region, e.g., $\xi_c=1.5$$-$$3.8$
in the three supershells. Therefore, if one applies the $I_{100}$-excess method
for deriving an accurate molecular content, the correction factor should
be estimated too.
An alternative method for estimating molecular content is to use the
$\tau_{100}$ excess. In Paper I, we showed that, for the Galactic worm 
GW46.4+5.5, the $\tau_{100}$ excess had a very good correlation 
with $W_{\rm CO}$ but the $I_{100}$ excess did not.
For the three supershells, however, the difference between the two methods is
not as prominent as in GW46.4+5.5.
There are several reasons for this: First the $I_{100}$ excess
due to high infrared emissivity caused by, e.g., heating by OB stars, is rare at
high latitudes and, since the effect is local, it would not significantly 
affect the correlation between the $I_{100}$ excess and $W_{\rm CO}$ 
over a large area. Second our observations are biased because 
we observed only the peak excess positions and the surrounding area. 
If we had coverd a larger area, the
difference might have been more significant.

In conclusion we consider that the $\tau_{100}$-excess method, in general,
is a more reliable method for estimating the molecular content than the
$I_{100}$-excess method. It might be worthwhile to compare the method with
other suggested methods in regions with fully sampled HI and CO data.

\acknowledgements
This work has been supported in part by the Basic Science Research Institute
Program, Ministry of Education, 1997, project BSRI-97-5408.

\end{document}